\documentclass[namedreferences]{SolarPhysics}
\usepackage[optionalrh]{spr-sola-addons} 

\usepackage[pdfborder={0 0 0},urlcolor=blue,breaklinks]{hyperref}
\usepackage{epsfig}          
\usepackage{graphicx}        
\usepackage{courier}         
\usepackage{amssymb}        
\usepackage{color}           

\newcommand{\be}{\begin{equation}}
\newcommand{\ee}{\end{equation}}
\newcommand{\beq}{\begin{eqnarray}}
\newcommand{\eeq}{\end{eqnarray}}

\usepackage{stdclsdv}
\usepackage{enumerate}

\newcommand{\aap}{    {\it Astron. Astrophys.}}

\newcommand{\apj}{    {\it Astrophys. J.}}

\newcommand{\grl}{    {\it Geophys. Res. Lett.}}

\newcommand{\jgr}{    {\it J. Geophys. Res.}}

\newcommand{\solphys}{{\it Solar Phys.}}

\begin{document}
\begin{article}
\begin{opening}

\title{The Electron Firehose and Ordinary-Mode Instabilities in Space Plasmas}
\author{M. \surname{Lazar}$^{1,2}$ $\sep$S. \surname{Poedts}$^{1}$
$\sep$R. \surname{Schlickeiser}$^{2}$ $\sep$D.
\surname{Ibscher}$^{2}$}

\runningauthor{M. Lazar \emph{et al.}} \runningtitle{Firehose
\emph{versus} Ordinary-Mode Instabilities in Space Plasmas}

\institute{$^1$Center for Mathematical Plasma Astrophysics, K.U.
Leuven, Celestijnenlaan 200B, 3001 Leuven, Belgium\\
$^2$Institut f\"ur Theoretische Physik, Lehrstuhl IV: Weltraum- und
Astrophysik, Ruhr-Universit\"at Bochum, D-44780 Bochum, Germany,
\email{mlazar@tp4.rub.de}.}
\date{Received ; accepted }

\begin{abstract}
The selfgenerated wave fluctuations are particularly interesting in
the solar wind and magnetospheric plasmas, where Coulomb collisions
are rare and cannot explain the observed states of
quasi-equilibrium. Linear theory predicts that the firehose and the
ordinary-mode instabilities can develop under the same conditions,
confusing the role of these instabilities in conditioning the
space-plasma properties. The hierarchy of these two instabilities is
reconsidered here for nonstreaming plasmas with an electron
temperature anisotropy $T_\parallel > T_\perp$, where $\parallel$
and $\perp$ denote directions with respect to the local mean
magnetic field. In addition to the previous comparative analysis,
here the entire 3D wave-vector spectrum of the competing
instabilities is investigated, paying particular attention to the
oblique firehose instability and the relatively poorly known
ordinary-mode instability. Results show a dominance of the oblique
firehose instability with a threshold lower than the parallel
firehose instability and lower than the ordinary-mode instability.
For larger anisotropies, the ordinary mode can grow faster, with
maximum growth rates exceeding the ones of the oblique firehose
instability. In contrast to previous studies that claimed a possible
activity of the ordinary-mode in the small $\beta [< 1]$ regimes,
here it is rigorously shown that only the large $\beta [> 1]$
regimes are susceptible to these instabilities.

\end{abstract}

\keywords{Corona; Flares, Dynamics; Solar wind; Instabilities;
Waves, Plasma}

\end{opening}

\section{Introduction}\label{Introduction}

Because space-plasmas are hot and weakly collisional, large
deviations from thermal equilibrium are expected to be observed even
for periods of a quiet Sun. But this is not confirmed by the
\emph{in-situ} measurements of the particle velocity distributions,
which show a relatively small temperature anisotropy for both
species: electrons and ions (or protons) (for a review, see
\opencite{ma06}). Kinetic instabilities have been found to be very
efficient to reduce the free energy and scatter particles,
preventing an increase of the temperature anisotropy, \emph{e.g.} an
increase of temperature in the direction of a guiding magnetic field
is predicted by the adiabatic expansion
\cite{ga93,ga99,he06a,st08,ba09}).

Here we assume such an excess of parallel temperature, namely,
$T_\parallel > T_\perp$ (where $\parallel$ and $\perp$ denote
directions relative to the uniform magnetic field), which can drive
two distinct instabilities: the firehose instability (FHI) and the
ordinary-mode instability (OMI). The FHI has been extensively
studied (see, for instance, \opencite{ga93}, \opencite{li00},
\opencite{ga03}, \opencite{pa03}, \opencite{ca08}, \opencite{la09}
and references therein), providing a quite precise picture of its
potential role in temperature isotropization and energy dissipation
in the solar wind, flares, and coronal mass ejections. The OMI is
less known in this context. This is an aperiodic instability, driven
by the velocity anisotropy of plasma particles, \emph{e.g.} streams,
temperature anisotropy, in the direction of lower energy. These
features suggest a close kinship with the Weibel instability
originally described by \inlinecite{we59} and \inlinecite{fr59} in
field-free plasmas. In the presence of a uniform magnetic field
$[{\bf B}_0]$, an excess of parallel temperature $[ T_\parallel >
T_\perp ]$ may destabilize the ordinary-mode $[ \delta {\bf E} \perp
{\bf B}_0 ]$ in the perpendicular direction ${\bf k} \perp {\bf
B}_0$. Recently, the OMI has been reexamined \cite{ib12}, providing
an accurate characterization of the instability conditions, and
these results are invoked in the present analysis.

Recent investigations \cite{la09,la10} suggest a potential
competition between these instabilities at high frequencies, where
both the FHI and OMI are driven by electrons with $T_{{\rm
e},\parallel} > T_{{\rm e},\perp}$. Thus, for the relaxation of
sufficiently large anisotropies, linear dispersion theory predicts
maximum growth rates comparable to the proton gyrofrequency for the
parallel FHI, while growth rates of the OMI can be several orders of
magnitude larger (see Figures~\ref{f2} and \ref{f3}). On the other
hand, the anisotropy threshold of the FHI seems to be lower, giving
to this instability chances to develop, but only for small
anisotropies close to the threshold values. However, analytical
approximations proposed to describe the OMI solutions in the limit
of large wavelengths (larger than the electron gyroradius,
\emph{i.e.} $\lambda \equiv k^{-1} > \rho_{\rm e} \equiv u_{{\rm e},
\perp} /\Omega_{\rm e}$; \opencite{ha68}), may lead to unrealistic
estimations of the instability threshold (compare Figure~\ref{f4}
with the results of \opencite{la10}). A realistic analysis should
also include the oblique FHI $[ 0< \theta < \pi/2 ]$, which develops
faster than the parallel branch \cite{pa99, ga03, ca08}. At oblique
angles $[ |{\bf k}\cdot{\bf B}_0| /(kB_0) = |\cos \theta| \ne 1 ]$,
this instability exhibits two distinct branches. The first branch is
supplied by the propagating modes (with nonzero-frequency $\omega_r
\equiv \Re(\omega) \ne 0$), which are also present in the direction
parallel to the magnetic field. But the FHI seems to be dominated by
the second branch of non-propagating (or zero-frequency $\omega_r =
0$) modes, which occur only for oblique propagation $[ {\bf k}
\times {\bf B}_0 \ne 0 ]$.

\begin{figure}[h]
\centerline{
    \includegraphics[width=45mm]{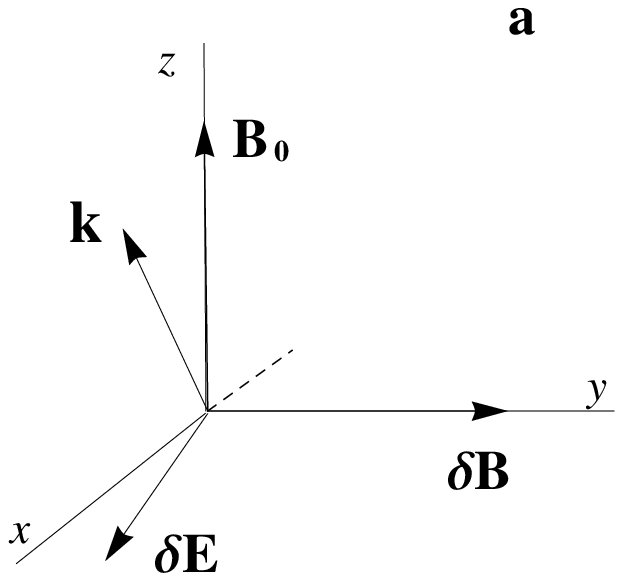}
    \includegraphics[width=45mm]{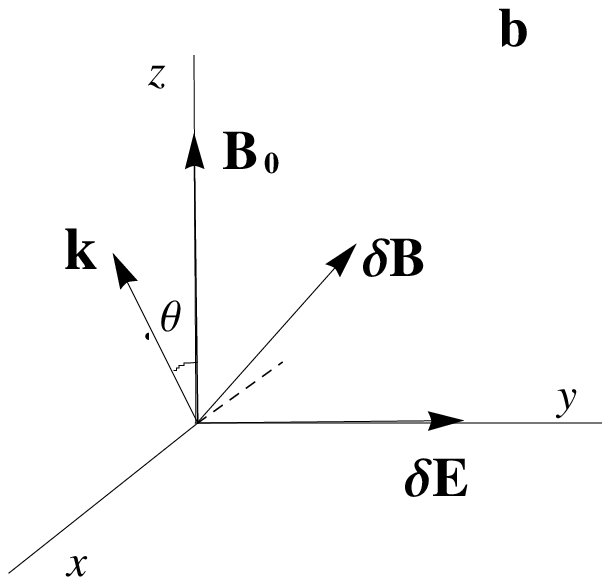}
    \includegraphics[width=45mm]{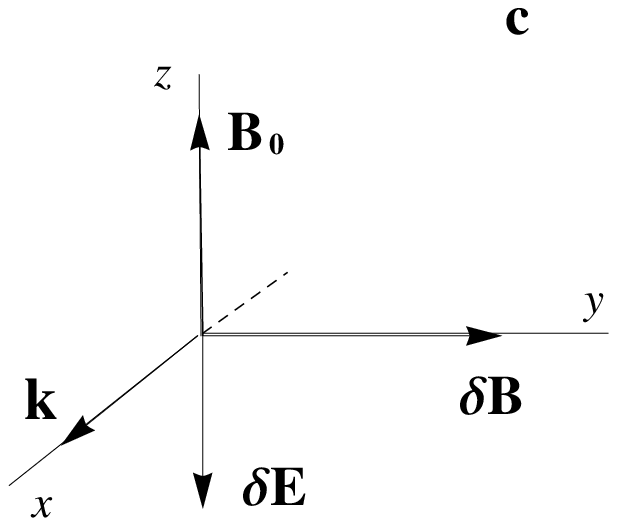}}
    \caption{The wave electric and magnetic-field orientation:
    (a) propagating firehose, (b) non-propagating firehose,
    (c) ordinary-mode (OMI)} \label{f1}%
\end{figure}

\begin{figure}[h]
\centerline{\includegraphics[width=120mm]{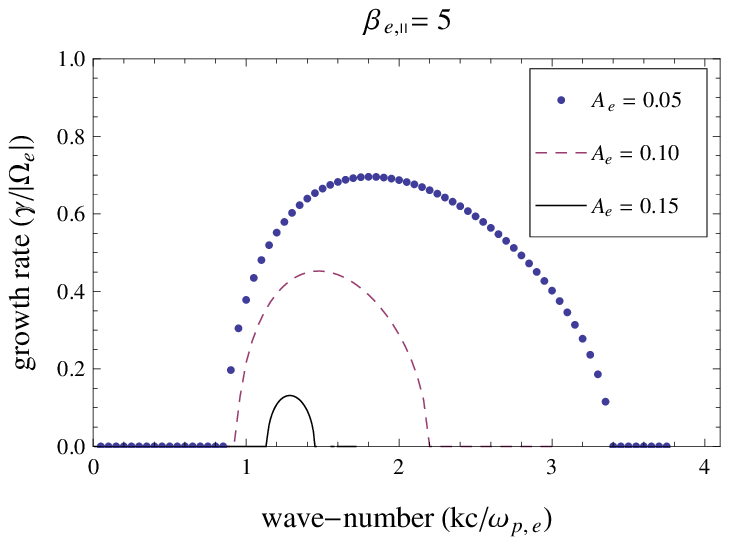}}
\centerline{\includegraphics[width=120mm]{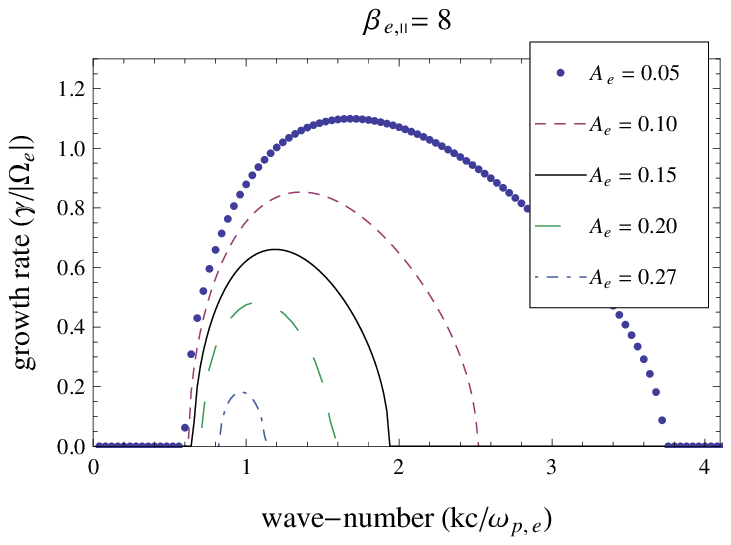}}
    \caption{Growth rates of the OMI for sufficiently large
    temperature anisotropies above the threshold (see Table \ref{tab1}):
    $A_{\rm e} < 0.155$ for $\beta_{{\rm e},\parallel} = 5$,
    and $A_{\rm e} < 0.276$ for $\beta_{{\rm e},\parallel} = 8$.} \label{f2}%
\end{figure}

For a clear picture, we propose to compare the growing modes
starting from the orientation of their wave-field vectors in
Figure~\ref{f1}. Thus, while the propagating FHI is a shear
transverse mode (Figure~\ref{f1} (a)), the non-propagating FHI
(Figure~\ref{f1} (b)) has a compressive component $\delta
B_\parallel = \delta B \sin \theta \ne 0$, which becomes dominant at
large angles of propagation, \emph{i.e.} $|\delta B_\parallel|
> |\delta B_\perp|$. The ordinary-mode is linearly polarized, and
the orientation of the wave-field vectors is shown in
Figure~\ref{f1} (c). To correlate and extract maximum of information
from recent numerical simulations, we keep their settings choosing
the coordinate system such that both ${\bf B}_0$ and the wave-vector
$[{\bf k}]$ lie in the $x-z$ plane \cite{ga03}. These simulations
clearly demonstrate that i)~the fluctuating fields during the growth
phase are due to a zero-frequency mode, and ii)~throughout the
growth, saturation, and subsequent decay of the fields, the dominant
component of the fluctuating magnetic field is $\delta B_y$,
satisfying $|\delta B_x|^2 \ll |\delta B_z|^2 \ll |\delta B_y|^2$.
According to Figure~\ref{f1}, a major fluctuating magnetic component
$\delta B_y$ cannot be attributed to the non-propagating FHI. It
could be associated with the propagating FHI mode, but growth rates
of this oscillatory mode are much lower, and it is not confirmed by
simulations. Instead, the OMI (Figure~\ref{f1} (c)) could offer a
plausible explanation, as it drives a purely growing
(non-propagating) magnetic field fluctuation $\delta B_y$. However,
it is not yet demonstrated whether this instability can arise and
compete or not with the oblique FHIs.

The oscillatory (propagating) modes have been extensively
investigated to establish their drivers as well as their effects in
different scenarios in space plasmas (\emph{e.g.}, see textbook by
\inlinecite{ga93} and references therein). However, recent studies
pay special attention to non-propagating wave instabilities,
\emph{e.g.} the compressive mirror and FHIs, and the OMI, which
appear to provide better explanations for the observed distributions
of data in the solar wind and terrestrial magnetosphere
\cite{he06a,st08,ba09}. The \emph{in-situ} measurements of
interplanetary magnetic-field fluctuations have indeed confirmed a
nearly perpendicular wave-vector $[k_\perp \gg k_\parallel]$ power
distribution \cite{sa12}. This article is therefore intended to
present refined comparative analysis of these instabilities, using
recent results for a rigorous characterization of the OMI
\cite{ib12}, and covering the entire 3D wave-vector spectrum of the
FHI.

We assume an homogeneous plasma system, largely extended and
immersed into an uniform magnetic field $[{\bf B_0}]$. The
unperturbed particle velocity distribution is anisotropic, with an
excess of temperature in the direction of the magnetic field, i.e.,
$T_\parallel > T_\perp$. To model this anisotropy, we use a
gyrotropic bi-Maxwellian distribution function
\be F_{\rm a} = {1 \over (2\pi)^{3/2} u_{{\rm a},\parallel} u_{{\rm
a},\perp}^2} \, \exp \left(-{v_{\parallel}^2 \over 2 u_{{\rm
a},\parallel}^2} - {v_{\perp}^2 \over 2 u_{{\rm a},\perp}^2}
\right), \label{e1} \ee
where $v_{\parallel}$ and $v_{\perp}$ are, respectively, the
particle velocity components parallel and perpendicular to ${\bf
B}_0$, and $u_{{\rm a},\parallel} = (k_B T_{{\rm a},\parallel}
/m_{\rm a})^{1/2}$ and $u_{{\rm a},\perp} = (k_BT_{\perp, {\rm a}}
/m_{\rm a})^{1/2}$ are the corresponding thermal velocities for the
plasma particles of types a (a $=$ e  for electrons, a $=$ i for
ions, and a $=$ p for protons). The stability analysis of a hot
collisionless plasma is based on the linearized Vlasov--Maxwell
equations. Here we investigate the unstable wave-mode solutions
driven by the anisotropy of the electron temperature, namely, an
excess of the parallel temperature, \emph{i.e.} $T_{{\rm
e},\parallel} > T_{{\rm e},\perp}$. These modes are the
ordinary-mode, which propagates perpendicular to the magnetic field
(${\bf k} \perp {\bf B}_0$), and the firehose mode, which propagates
parallel or obliquely to the magnetic field $[{\bf k} \cdot {\bf
B}_0 \ne 0]$. These are schematically shown in Figure~\ref{f1}. In
contrast to parallel propagation, where the electrostatic and
electromagnetic modes are decoupled and their theory is relatively
simple, at oblique or perpendicular propagation the dispersion
relations are complicated, and an accurate characterization of their
solutions is possible only numerically.


\section{The Ordinary-Mode Instability (OMI)}%


For anisotropic plasmas modeled by the distribution function in
Equation (\ref{e1}), ordinary modes are described by the dispersion
relation \cite{ib12}
\begin{equation}
{\omega^2 - k^2c^2\over \omega_{\rm p, e}^2}=  1 +{2 \over A_{\rm
e}} e^{-{k^2 u_{{\rm e},\perp}^2 \over \Omega_{\rm e}^2}}
\sum_{n=1}^{\infty} \frac{n^2 \Omega_{\rm e}^2 \; I_n({k^2 u_{{\rm
e},\perp}^2 / \Omega_{\rm e}^2})}{\omega^2 -n^2\Omega_{\rm e}^2},
\label{e2}
\end{equation}
where $I_n$ are the modified Bessel functions of the first kind, and
$A = T_\perp /T_\parallel$ corresponds to the temperature
anisotropy. Because the ions (protons) are much heavier than the
electrons ($m_{\rm i} > m_{\rm p} \gg m_{\rm e}$), their effects can
be neglected at sufficiently high frequencies, and we can assume
that they are rigid or isotropically distributed ($T_\parallel =
T_\perp$).
\begin{table}[h]
\caption{The anisotropy thresholds ($\gamma_{\rm m} =0$) from
Equation~(\ref{e3}) for different values of $\beta_{{\rm e},
\parallel}$.} \label{tab1}
\begin{tabular}{c c c c c c c c c c}\hline
$\beta_{{\rm e},\parallel}$ & 2.5 & 5 & 8 & 16 & 30 & 60 & 100 & 500 & 1000\\
\hline $(A_{\rm e})_{\rm threshold}$& 0.014 & 0.155 & 0.276 & 0.442
& 0.563 & 0.660 & 0.711 & 0.802 & 0.823 \\
\hline
\end{tabular}
\end{table}
The marginal instability threshold can be calculated numerically
solving the dispersion relation in Equation (\ref{e2}) for a maximum
growth rate $\gamma_{\rm m} \to 0$ \cite{ib12}. Here, for
simplicity, we use the analytical form
\begin{equation}
A_{\rm e} \leqslant 1-\frac{2 a(\beta_{{\rm e}, \parallel})}
{\beta_{{\rm e}, \parallel}}, \;\; {\rm with} \;\;\; a(\beta_{{\rm
e} \parallel}) = \ln \left(w \beta_{{\rm e},\parallel} \over 2
\right) + {1 \over w} I_{0} \left[\ln \left(w \beta_{{\rm
e},\parallel} \over 2 \right) \right], \label{e3}
\end{equation}
which, for a value of the fitting parameter $w = 0.9$, provides an
accurate fit with the exact numerical threshold \cite{ib12}. Values
of the anisotropy threshold are given in Table~\ref{tab1} for
different values of $\beta_{e\parallel} \equiv 8\pi nk_B T_{{\rm
e},\parallel}/B_0^2$, and are also displayed in Figure~\ref{f4}
(solid line).

Notice that the existence of the OMI is clearly limited to
sufficiently large $\beta_{{\rm e}, \parallel} [ > 1]$ regimes. The
exact instability threshold in Equation~(\ref{e3}) is markedly
different from the instability condition
\be A_{\rm e}  \leqslant {2 \over 3} \left(1- {1 \over \beta_{{\rm
e},\parallel}} \right)^2, \label{e5} \ee
obtained by \inlinecite{ha68} in the limit of a small argument of
the Bessel function $[I_n(x <1)]$.

\begin{figure}[h]
\centerline{
    \includegraphics[width=100mm]{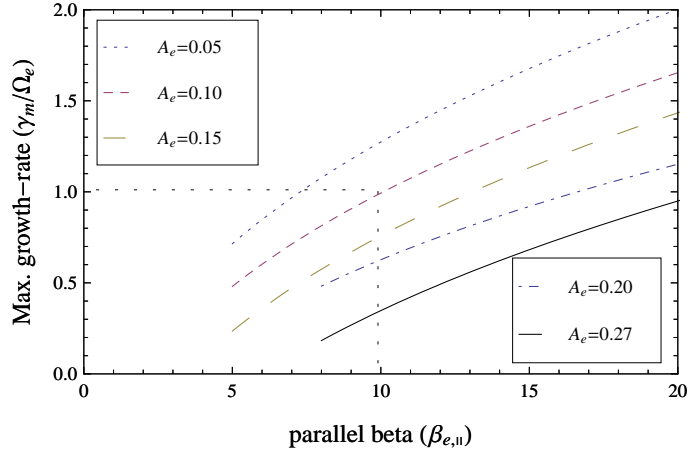}}
    \caption{Maximum growth rate of the ordinary-mode instability
    as a function of $\beta_{{\rm e},\parallel}$ for different values of
    the temperature anisotropy: $A_e = 0.05$, 0.10, 0.15, 0.20 and
    0.27.} \label{f3}%
\end{figure}

Growth rates $[\gamma = \omega_i]$ of the OMI are displayed in
Figure~\ref{f3}. These are calculated numerically for conditions
typically encountered at different altitudes [0.3 -- 1~AU] in the
solar wind. At the saturation, maximum growth rates $[\gamma_{\rm
m}]$ depend only on the plasma beta $[\beta_{{\rm e},\parallel}]$
and the temperature anisotropy $[A_{\rm e}]$. For sufficiently large
anisotropies, $\gamma_{\rm m}$ can approach or exceed $|\Omega_{\rm
e}|$, see the right panel in Figure~\ref{f3}. Recent estimates have
restricted \cite{ib12}
\be \gamma_{\rm m} \leqslant \left|k^{2}c^{2}+ \frac{\omega_{\rm
p,e}^{2}} {A_{\rm e}}\left\{A_{\rm e}-1+I_0 \left(g
\right)\exp[-g]\right\}\right|^{1/2}, \label{e6} \ee
where
\be g = \frac{k^2u_{{\rm e},\perp}^2}{\Omega_{\rm e}^2} = {k^2c^2
\over \omega_{\rm p,e}^2} {A_{\rm e} \beta_{{\rm e},\parallel}\over
2}, \label{e7} \ee
but this limit is function of the wavenumber value, and cannot
provide general constraints depending only on the plasma parameters.
Here, in Figure~\ref{f3}, we derive numerically the exact values of
the maximum growth rates $[\gamma_{\rm m}]$ function of $\beta_{{\rm
e},\parallel}$, and for different values of the temperature
anisotropy $[A_{\rm e} < 1]$. The OMI is enhanced by the temperature
anisotropy $[T_{{\rm e},\parallel} > T_{{\rm e},\perp}]$, but it is
inhibited by the stationary magnetic field ($\beta_{{\rm
e},\parallel} \sim B_0^{-2}$).

\section{Interplay with the Oblique FHI}

Previous studies have shown that the parallel FHI cannot compete
with the OMI, which is much faster \cite{la09, la10}. Here we give
particular attention to the oblique FHI, which grows much faster
than the parallel FHI. The predictions of the dispersion theory and
numerical confirmations are clear in this case: the parallel
firehose instability admits maximum growth rates close to, but less
than, $\Omega_{\rm p}$ \cite{ga85}, whereas the oblique firehose
instability can reach maximum growth rates two or three orders of
magnitudes larger, \emph{e.g.} $\Omega_{\rm p} \ll \gamma_{\rm m} <
|\Omega_{\rm e}|$ \cite{li00,ga03,ca08}. Moreover, the anisotropy
threshold of the oblique firehose instability is also lower
\cite{pa99,he06a,st08}.

At oblique propagation, the electron-firehose instability splits
into two different branches, and both develop much faster than the
parallel firehose instability. The first branch is the continuation
of the parallel firehose instability and is supplied by the
propagating (nonzero-frequency) modes. In the second branch, the
instability is purely growing (or aperiodic), and is usually called
the non-propagating FHI. Now, it is worth mentioning that a powerful
numerical resolution of the Vlasov--Maxwell dispersion relations
clearly shows that i)~the existence of propagating modes at large
angles ($\theta$ up to $70^o - 80^o$) (even for a small anisotropy
$T_{{\rm e},\parallel}/T_{{\rm e},\perp} \gtrsim 2$; ii)~the maximum
growth rates are reached for large angles ($\theta > 45^o$); and
iii)~the nonpropagating FHI is largely dominant at these
inclinations \cite{ca08}.

Looking to the field properties in Figure~\ref{f1}, we can add
further distinctions between these two branches. Like its
proton-driven counterpart, the propagating electron firehose mode
(Figure~\ref{f1} (a)) is a shear (torsional) transverse wave that
twists magnetic field lines relative to one another but does not
compress \cite{sw03, ga03}. This mode is nonresonant with electrons
but resonant with ions \cite{ga93,pa99}, enabling the transfer of
energy from electrons to protons, and thus supporting the
transit-time damping scenario. On the other hand, the nonpropagating
$[\omega_{\rm r} = 0]$ mode (Figure~\ref{f1} (b)) has a compressive
component, $\delta B_\parallel \ne 0$, parallel to the mean magnetic
field. When this is small, that is, when $|\delta B_\parallel| <
|\delta B_\perp|$ [small angles $\theta$], the instability is
predominantly transverse and cyclotron resonant with electrons, and
it can therefore play an important role in the relaxation of their
anisotropy \cite{ga03}. When the parallel component is large, that
is when $|\delta B_\parallel| > |\delta B_\perp|$, the instability
is predominantly compressive, like the electron mirror instability,
which also has $\omega_{\rm r} = 0$, but is driven by an opposite
anisotropy $T_{{\rm e},\perp} > T_{{\rm e},\parallel}$
\cite{po02,ga06}. Notice that the thresholds of these two
instabilities provide the best fit to the observed limits of the
temperature anisotropy in space plasmas \cite{he06a,st08}. In this
case, the wave magnetic field not only rotates but changes its
magnitude as well. Landau damping of this nonpropagating mode can be
very efficient at scattering electrons in phase space \cite{ga06}.

\begin{figure}[h]
\centerline{\includegraphics[width=100mm]{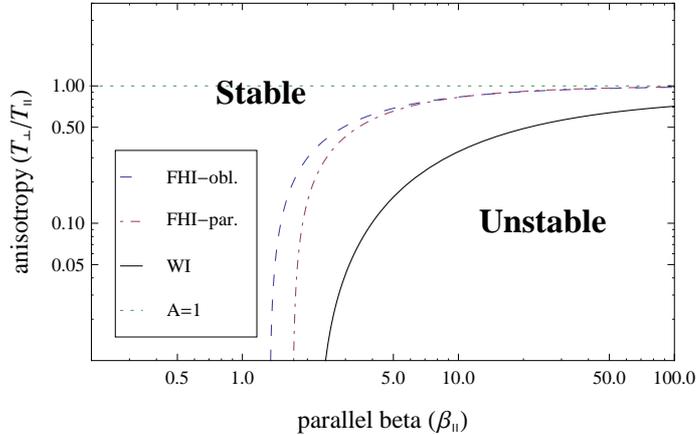}}
    \caption{The instability thresholds: comparison of the parallel
    FHI (dash-dotted line), the non-propagating FHI (long-dashed line)
    and the OMI (solid line).} \label{f4}%
\end{figure}

In Figure~\ref{f4}, we compare the anisotropy thresholds of these
two instabilities. The propagating FHI corresponds to the
dash-dotted line, and the non-propagating FHI corresponds to the
dashed line. The exact thresholds are derived numerically for
finite, but very small values of the growth rate, e.g., $\gamma_{\rm
m} = 10^{-2} \Omega_{\rm p}$, close to the marginal condition of
stability $[\gamma_{\rm m} = 0]$. The isolines of constant growth
rate derived from linear theory are usually fitted with
\be {T_{{\rm e},\perp} \over T_{{\rm e},\parallel}} = 1- {S \over
\beta_{{\rm e},\parallel}^\alpha}. \label{e8} \ee
Values of the fitting parameters [$S$ and $\alpha$] derived for a
$\gamma_{\rm m} = 10^{-2} \Omega_{\rm p}$ are shown in
Table~\ref{tab2} (after \opencite{ga03}).
\begin{table}[h]
\caption{Values of the fitting parameters $S$ and $\alpha$ in
Equation~(\ref{e8}).} \label{tab2}
\begin{tabular}{c c c} \hline
($\gamma_{\rm m} = 10^{-2} \Omega_{\rm p}$) & Propagating FHI & Nonpropagating FHI \\
\hline $S$ & 1.70 & 1.29 \\ $\alpha$ & 0.99 & 0.88 \\ \hline
\end{tabular}
\end{table}

For comparison, the threshold of the OMI as given by
Equation~(\ref{e3}) is illustrated in Figure~\ref{f4} with the solid
line. We can also fit to power laws of the form in
Equation~(\ref{e8}), but Equation~(\ref{e3}) is the best analytical
description of the instability threshold (see the analysis by
\opencite{ib12}). Thresholds of the FHI are lower than those of the
OMI, but the maximum growth rates of the latter increase much faster
with $\beta_{{\rm e},\parallel}$ (\emph{e.g.} see Figure~\ref{f3}),
leading to values larger than $|\Omega_{\rm e}|$, which is the limit
derived numerically for the nonpropagating FHI \cite{ga03,ca08}.
This supports our result that for sufficiently large anisotropies,
\emph{e.g.} $T_{{\rm e},\parallel}/T_{{\rm e},\perp} > 10$, or
sufficiently large $\beta_{{\rm e},\parallel} > 10$, the OMI is
faster than the FHI and can manifest itself as the principal
mechanism of relaxation. Larger values of $\beta_{{\rm
e},\parallel}$ means less intense magnetic fields $[B_0]$, or/and
more dense and hotter plasma populations, so that these conditions
can be encountered at different altitudes in the solar wind.

\section{Discussion and Conclusions}

We have used results of the linear Vlasov-Maxwell theory to compare
the properties of two growing modes driven by $T_{{\rm e},
\parallel} > T_{{\rm e},\perp}$: the FHI and the OMI. In addition to previous
analyses, here the entire 3D wave-vector spectrum of the competing
instabilities is examined, paying particular attention to the
non-propagating FHI. Before drawing the main conclusions of our
article, we examine the conditions for the initiation of these
instabilities in recent numerical simulations, and in solar-wind
observations. Numerical studies have reported on linear and
quasi-linear developments of the electron FHI in PIC simulations
\cite{me02,ga03,ca08} and test particle simulations \cite{pa03}. The
conditions implemented in these experiments are typical for the
solar wind, or for the impulsive solar flares. To shed more light on
these results, we refer again to the wave-field vectors
representation in Figure~\ref{f1}. We keep the settings from the
simulations choosing the coordinate system such that both ${\bf
B}_0$ and the wave-vector $[{\bf k}]$ lie in the $x-z$ plane
\cite{ga03,ca08}.

Numerical simulations demonstrated that \romannumeral 1)~the
fluctuating fields during the growth phase are due to a
nonpropagating (zero-frequency) mode, and \romannumeral
2)~throughout the growth phase, the saturation phase, and the
subsequent decay phase of the fields, the dominant component of the
fluctuating magnetic field is $|\delta {\bf B}| =\delta B_y$,
satisfying $|\delta B_x|^2 \ll |\delta B_z|^2 \ll |\delta B_y|^2$.
(In the article by \inlinecite{ca08}, the cartesian system is
rotated, such that the dominant component is $\delta {\bf B}_z$, but
this corresponds to the same component perpendicular to both ${\bf
k}$ and ${\bf B}_0$.) This component has been attributed to the
nonpropagating FHI, which can extend to quasi-perpendicular
directions and grows faster than the propagating FHI. However, the
orientation of the wave-field vectors in Figure~\ref{f1}
demonstrates that a major fluctuating magnetic component $\delta
B_y$ cannot be attributed to the non-propagating FHI. Moreover, at
large, quasi-perpendicular angles $[\theta]$, the nonpropagating
mode is highly compressional and exhibits large parallel
fluctuations $\delta B_{\parallel} = \delta B_z$. Instead,
magnetic-field fluctuations along the $y-$axis direction,
perpendicular to ${\bf B}_0$, can be driven by the propagating FHI,
but this is time-oscillatory (nonzero-frequency) and less apparent
in the early stage of the simulation. The OMI is also a plausible
candidate as it is a zero-frequency mode of comparable wavelength
(\emph{i.e.} $\approx c/\omega_{\rm p,e}$, that is the electron
skin-depth, see Figure~\ref{f2} above and Figure~3 of
\inlinecite{ca08}). 

Now we apply the results of this article to check if the OMI is fast
enough to develop and to compete with the FHI under the same
conditions of the two sets of PIC simulations \cite{ga03,ca08}. If
we consider the representative run F-257 from the 1D PIC simulations
of \inlinecite{ga03}, with initial parameters $\beta_{{\rm
e},\parallel} = 7.7$ and $A_{\rm e}= T_{{\rm e},\perp}/ T_{{\rm
e},\parallel} =0.46$, the OMI cannot set in, because the instability
threshold in Equation~(\ref{e3}) requires a temperature anisotropy
$A_{\rm e} < 0.27$. The same result is obtained if we check all the
other 1D runs (F-250, F-252, and F-255), or the 2D runs of
\inlinecite{ca08}, concluding that none of these PIC simulations are
relevant for the OMI. The instability cannot be initiated when
$A_{\rm e}$ is above the threshold.

Large temperature anisotropies $[T_{{\rm e},\parallel} \gg T_{{\rm
e},\perp}]$ are believed to arise in flares and other violent
phenomena in the solar wind, such as the co-rotating forward and
reversed shocks in coronal mass ejections. Under these conditions,
the parallel $\beta$ can cover a wide interval of values ($0.01<
\beta_{{\rm e},\parallel} < 100$), and the OMI is much easier
excited, offering plausible explanations for the observed emissions
and suprathermal populations. For instance, for a moderate value of
$\beta_{{\rm e},\parallel}=8$, the instability threshold in
Equation~(\ref{e3}) requires a temperature anisotropy $A_{\rm e} <
0.28$. This is equivalent with a minimum $T_{{\rm e},\parallel}
/T_{{\rm e},\perp}> 3.6$ that is well below the large temperature
anisotropies $[T_{{\rm e},\parallel} /T_{{\rm e},\perp}> 10]$
predicted by the observations in flares. Such scenarios have been
implemented in numerical simulations, but only the parallel FHI has
been examined \cite{me02,pa03}, without indications from directions
perpendicular to the magnetic field. More information can be
extracted from the solar-wind observations of their proton-driven
counterparts. Measurements of the magnetic-field fluctuations show
an enhanced magnetic compressibility along the mirror instability
threshold (at $T_\perp >T_\parallel$ and large $\beta_\parallel
>1$), but small elsewhere \cite{ba09}. This is consistent with our
results, which suggest non-compressive instability constraints for
the opposite anisotropies $T_\perp < T_\parallel$. However, the
magnetic compressibility is increasing with increasing
$\beta_\parallel$ (Figure~1 of \inlinecite{ba09}), reaching values
comparable with the mirror instability if $\beta_\parallel > 8$ is
large enough, and indicating traces of the quasi-perpendicular
compressive modes.

\begin{table}[h]
\caption{Properties of the wave instabilities driven by $T_\parallel
> T_\perp$.} \label{tab3}
\begin{tabular}{c c c c c}\hline
Instability & Growth rate & Threshold & Propagation & Fluctuating fields \\
\hline OMI, $\omega_{\rm r} = 0$ & $\gamma_{\rm m} > |\Omega_{\rm e}|$ & High & $\theta =\pi / 2$ & $\delta{\bf E} \parallel {\bf B}_0$\\
FHI, $\omega_{\rm r} \ne 0$  & $\gamma_{\rm m} \ll |\Omega_{\rm e}|$
& Low & $0 \leqslant \theta < \pi / 2$ & $\delta
{\bf B} \perp {\bf B}_0$\\
FHI, $\omega_{\rm r} = 0$ & $\gamma_{\rm m} < |\Omega_{\rm e}|$ & Lower & $0 < \theta < \pi / 2$ & $\delta{\bf B}_\parallel \ne 0$\\
\hline
\end{tabular}
\end{table}

In conclusion, the comparative analysis of the wave instabilities
driven in a magnetized plasma by an excess of parallel temperature
predicts a dominance of the non-propagating FHI with a threshold
lower than that of the parallel FHI and lower than that of the OMI.
However, for larger anisotropies, the ordinary mode can develop
faster with maximum growth rates exceeding that of the
non-propagating FHI. The properties of the competing wave
instabilities have been summarized in Table~\ref{tab3}. In contrast
to previous studies that claimed a possible activity of the OMI in
the small $\beta < 1$ regimes, here it is rigorously shown that only
the large $\beta> 1$ regimes are susceptible to these instabilities.
The existing numerical simulations are dedicated to the FHI, and
cannot offer information about the OMI because the settings are far
below the threshold condition. Numerical simulations confirm the
predictions of the dispersion theory that the fastest growing mode
is aperiodic. However, according to our analysis, the dominant
component of the fluctuating magnetic field indicated by the
simulations (perpendicular to ${\bf B}_0$) cannot be attributed to
the non-propagating FHI, but to the OMI, or, eventually, to the
propagating FHI, although such a time-oscillatory mode is not
reported in the early linear stage of the simulations. Motivated by
these results, the next numerical investigations should consider
supplementary conditions for the OMI to develop, and examine the
interplay with the FHI in both stages of linear growth and nonlinear
saturation.

\begin{acks}

ML acknowledges financial support from the EU Commission and
Research Foundation Flanders (FWO) as FWO Pegasus Marie Curie Fellow
(grant 1.2.070.13). The authors acknowledge support from the
Ruhr-Universit\"at Bochum, the Deutsche Forschungsgemeinschaft
(DFG), grant Schl 201/21-1, and by the Katholieke Universiteit
Leuven. These results were obtained in the framework of the projects
GOA/2009-009 (KU Leuven), G.0729.11 (FWO-Vlaanderen) and C~90347
(ESA Prodex 9). The research leading to these results has also
received funding from the European Commission's Seventh Framework
Programme (FP7/2007-2013) under the grant agreements SOLSPANET
(project n° 269299,
\href{http://www.solspanet.eu/solspanet}{www.solspanet.eu}),
SPACECAST (project n° 262468,
\href{http://fp7-spacecast.eu}{fp7-spacecast.eu}), eHeroes (project
n° 284461,
\href{http://soteria-space.eu/eheroes/html/}{www.eheroes.eu}) and
SWIFF (project n° 263340,
\href{http://www.swiff.eu/}{www.swiff.eu}).

\end{acks}



\end{article}

\begin{thebibliography}{}

\bibitem[\protect\citeauthoryear{Bale {\it et al.}}{2009}]{ba09} Bale,
S., Kasper, J.C., Howes, G.G., Quataert, E., Salem, E., Sundkvist,
D.: 2009, Magnetic fluctuation power near proton temperature
anisotropy instability thresholds in the solar wind. \emph{Phys.
Rev. Lett.} \textbf{103}, 211101.
\href{http://prl.aps.org/abstract/PRL/v103/i21/e211101}{doi:10.1103/PhysRevLett.103.211101}.

\bibitem[\protect\citeauthoryear{Camporeale and Burgess}{2008}]{ca08}
Camporeale, E., Burgess, D.: 2008, Electron firehose instability:
Kinetic linear theory and two-dimensional particle-in-cell
simulations. \jgr{} {\bf 113}, A07107.
\href{http://onlinelibrary.wiley.com/doi/10.1029/2008JA013043/abstract}{doi:10.1029/2008JA013043}.

\bibitem[\protect\citeauthoryear{Fried}{1959}]{fr59} Fried, B.D.:
1959, Mechanism for instability of transverse plasma waves. {\it
Phys. Fluids} {\bf 2}, 337.
\href{http://pof.aip.org/resource/1/pfldas/v2/i3/p337_s1?isAuthorized=no}{doi:10.1063/1.1705933}.


\bibitem[\protect\citeauthoryear{Gary}{1993}]{ga93} Gary, S.P.: 1993,
{\it Theory of Space Plasma Microinstabilities}, University Press,
Cambridge.

\bibitem[\protect\citeauthoryear{Gary and Karimabadi}{2006}]{ga06} Gary,
S.P., Karimabadi, H.: 2006, Linear theory of electron temperature
anisotropy instabilities: Whistler, mirror, and Weibel. \jgr{} {\bf
111}, A11224.
\href{http://onlinelibrary.wiley.com/doi/10.1029/2006JA011764/abstract}{doi:10.1029/2006JA011764}.

\bibitem[\protect\citeauthoryear{Gary and Madland}{1985}]{ga85} Gary,
S.P., Madland, D.: 1985, Electromagnetic electron temperature
anisotropy instabilities. \jgr{} {\bf 90}, 7607-7610.
\href{http://onlinelibrary.wiley.com/doi/10.1029/JA090iA08p07607/abstract}{doi:10.1029/JA090iA08p07607}.

\bibitem[\protect\citeauthoryear{Gary {\it et al.}}{1999}]{ga99} Gary,
S.P., Neagu, E., Skoug, R.M., Goldstein. B.~E.: 1999, Solar wind
electrons: Parametric constraints. \jgr{} {\bf 104}, 19843-19849.
\href{http://onlinelibrary.wiley.com/doi/10.1029/1999JA900244/abstract}{doi:10.1029/1999JA900244}.

\bibitem[\protect\citeauthoryear{Gary and Nishimura}{2003}]{ga03} Gary,
S.P., Nishimura, K.: 2003, Resonant electron firehose instability:
Particle-in-cell simulations. {\it Phys. Plasmas} {\bf 10},
3571-3576.
\href{http://pop.aip.org/resource/1/phpaen/v10/i9/p3571_s1?isAuthorized=no}{doi:10.1063/1.1590982}.

\bibitem[\protect\citeauthoryear{Hamasaki}{1968}]{ha68} Hamasaki, S.:
1968, Electromagnetic microinstabilities of plasmas in a uniform
magnetic induction. {\it Phys. Fluids} {\bf 11}, 2724-2727.
\href{http://pof.aip.org/resource/1/pfldas/v11/i12/p2724_s1?isAuthorized=no}{doi:10.1063/1.1691879}.

\bibitem[\protect\citeauthoryear{Hellinger {\it et al.}}{2006}]{he06a} Hellinger,
P., Travnicek, P., Kasper J.C., Lazarus, A.J.: 2006, Solar wind
proton temperature anisotropy: Linear theory and WIND/SWE
observations. \grl{} {\bf 33}, L09101.
\href{http://onlinelibrary.wiley.com/doi/10.1029/2006GL025925/abstract}{doi:10.1029/2006GL025925}.

\bibitem[\protect\citeauthoryear{Ibscher, Lazar, and Schlickeiser}
{2012}]{ib12} Ibscher, D., Lazar, M., Schlickeiser, R.: 2012, On the
existence of Weibel instability in a magnetized plasma. II.
Perpendicular wave propagation: The ordinary mode. {\it Phys.
Plasmas} {\bf 19}, 072116.
\href{http://adsabs.harvard.edu/abs/2012PhPl...19g2116I}{doi:10.1063/1.4736992}.

\bibitem[\protect\citeauthoryear{Lazar and Poedts}{2009}]{la09} Lazar, M.,
Poedts, S.: 2009, Limits for the firehose instability in space
plasmas. \solphys{} {\bf 258}, 119-128.
\href{http://link.springer.com/article/10.1007\%2Fs11207-009-9405-y}{doi:10.1007/s11207-009-9405-y}.


\bibitem[\protect\citeauthoryear{Lazar, Schlickeiser, and Poedts}{2010}]
{la10} Lazar, M., Schlickeiser, R., Poedts, S.: 2010, Nonresonant
electromagnetic instabilities in space plasmas: interplay of Weibel
and firehose instabilities. In: \emph{AIP Conf. Proc.}
\textbf{1216}, 280-283.
\href{http://proceedings.aip.org/resource/2/apcpcs/1216/1/280_1?isAuthorized=no}{doi:10.1063/1.3395855}.

\bibitem[\protect\citeauthoryear{Li and Habbal}{2000}]{li00} Li, X., Habbal,
S.R.: 2000, Electron kinetic firehose instability. \jgr{} {\bf 105},
27377-27385.
\href{http://onlinelibrary.wiley.com/doi/10.1029/2000JA000063/abstract}{doi:10.1029/2000JA000063}.

\bibitem[\protect\citeauthoryear{Marsch}{2006}]{ma06} Marsch, E.:
2006, Kinetic physics of the solar corona and solar wind. {\it
Living Rev. Solar Phys.} {\bf 3}, 1.
\href{http://solarphysics.livingreviews.org/Articles/lrsp-2006-1/}{doi:10.12942/lrsp-2006-1}.

\bibitem[\protect\citeauthoryear{Messmer}{2002}]{me02} Messmer, P.:
2002, Temperature isotropization in solar flare plasmas due to the
electron firehose instability. \aap{} {\bf 382}, 301-311.
\href{http://www.aanda.org/articles/aa/abs/2002/04/aa1789/aa1789.html}{doi:10.1051/0004-6361:20011583}.

\bibitem[\protect\citeauthoryear{Paesold and Benz}{1999}]{pa99} Paesold,
G., Benz, A.O.: 1999, Electron firehose instability and acceleration
of electrons in solar flares. \aap{} {\bf 351}, 741-746.

\bibitem[\protect\citeauthoryear{Paesold and Benz}{2003}]{pa03} Paesold,
G., Benz, A.O.: 2003, Test particle simulation of the electron
firehose instability. \aap{} {\bf 401}, 711-720.
\href{http://www.aanda.org/articles/aa/abs/2003/14/aa3178/aa3178.html}{doi:10.1051/0004-6361:20030113}.

\bibitem[\protect\citeauthoryear{Pokhotelov {\it et al.}}{2002}]{po02} Pokhotelov,
O.A., Treumann, R.A., Sagdeev, R.Z., Balikhin, M.A., Onishchenko,
O.G., Pavlenko, V.P., Sandberg, I.: 2002, Linear theory of the
mirror instability in non-Maxwellian space plasmas. \jgr{} {\bf
107}, 1312.
\href{http://onlinelibrary.wiley.com/doi/10.1029/2001JA009125/abstract}{doi:10.1029/2001JA009125}.

\bibitem[\protect\citeauthoryear{Salem {\it et al.}}{2012}]{sa12} Salem, C.S.,
Howes, G.G., Sundkvist, D., Bale, S.D., Chaston, C.C., Chen, C.H.K.,
Mozer, F.S.: 2012, Identification of kinetic Alfv\'en wave
turbulence in the solar wind. \apj{} {\bf 745}, L9.
\href{http://iopscience.iop.org/2041-8205/745/1/L9/}{doi:10.1088/2041-8205/745/1/L9}.

\bibitem[\protect\citeauthoryear{Schlickeiser, Lazar, and Skoda}{2011}]{sc11}
Schlickeiser, R., Lazar, M., Skoda, T. 2011, Spontaneously growing,
weakly propagating, transverse fluctuations in anisotropic
magnetized thermal plasmas. {\it Phys. Plasmas} {\bf 18}, 012103.
\href{http://pop.aip.org/resource/1/phpaen/v18/i1/p012103_s1?isAuthorized=no}{doi:10.1063/1.3532787}.

\bibitem[\protect\citeauthoryear{Stverak {\it et al.}}{2008}]{st08} Stverak, S.,
Travnicek, P., Maksimovic, M., Marsch, E., Fazakerley, A.N., Scime,
E.E.: 2008, Electron temperature anisotropy constraints in the solar
wind. \jgr{} {\bf 113}, A03103.
\href{http://onlinelibrary.wiley.com/doi/10.1029/2007JA012733/abstract}{doi:10.1029/2007JA012733}.

\bibitem[\protect\citeauthoryear{Swanson}{2003}]{sw03}
Swanson, D.G.: 2003, {\it Plasma Waves, 2nd Edition}, IOP Publishing
Ltd., Bristol and Philadelphia.

\bibitem[\protect\citeauthoryear{Weibel}{1959}]{we59} Weibel, E.S.:
1959, Spontaneously growing transverse waves in a plasma due to an
anisotropic velocity distribution. {\it Phys. Rev. Lett.} {\bf 2},
83-84.
\href{http://prl.aps.org/abstract/PRL/v2/i3/p83_1}{doi:10.1103/PhysRevLett.2.83}.

\end{thebibliography}
\end{document}